\documentclass[11pt]{article}
\usepackage[a4paper, margin=1in]{geometry}
\usepackage[numbers]{natbib}
\usepackage{graphicx}
\usepackage{hyperref}
\usepackage{url}
\usepackage{booktabs}
\usepackage{pifont}
\usepackage{float}

\title{Interpretable LLMs for Credit Risk: A Systematic Review and Taxonomy}

\author{
  Muhammed Golec$^{1,2}$,
  Maha AlabdulJalil$^3$ \\
  \\
  $^1$School of Electronic Engineering and Computer Science, Queen Mary University of London, UK \\
  $^2$Institute for Data Science and Artificial Intelligence, Boğaziçi University, Turkey \\
  $^3$College of Science Computer Science Department, Kuwait University, Kuwait\\
}

\date{}

\begin{document}
\maketitle

\vspace{1em}
\begin{center}
\textit{This manuscript remains under review.}
\end{center}

\begin{abstract}
Large Language Models (LLM), which have developed in recent years, enable credit risk assessment through the analysis of financial texts such as analyst reports and corporate disclosures. This paper presents the first systematic review and taxonomy focusing on LLM-based approaches in credit risk estimation. We determined the basic model architectures by selecting 60 relevant papers published between 2020-2025 with the PRISMA research strategy. And we examined the data used for scenarios such as credit default prediction and risk analysis. Since the main focus of the paper is interpretability, we classify concepts such as explainability mechanisms, chain of thought prompts and natural language justifications for LLM-based credit models. 

The taxonomy organizes the literature under four main headings: model architectures, data types, explainability mechanisms and application areas. Based on this analysis, we highlight the main future trends and research gaps for LLM-based credit scoring systems. This paper aims to be a reference paper for artificial intelligence and financial researchers.

\end{abstract}

\section{Introduction}

Credit risk assessment is one of the important components in financial decision making, such as making investment decisions and determining whether to grant credit to an individual or institution \cite{nana2022game}. While traditional methods rely on structured data such as financial ratios and past payment information when assessing credit risk, in reality important information about the credit may not be available in an organized manner \cite{roeder2022data}. An example of this is that executive comments, economic news and analyst reports are generally in free text (unstructured).

Advanced LLM models such as GPT and FinBERT have great potential in financial applications with their high performance in extracting meaning from free text \cite{kang2025comparative,golec2025llm}. They can produce interpretable outputs by making sense of complex financial data and thus facilitate decision-making processes \cite{shobayo2024innovative}. With its high potential, this research area, LLM-Driven Credit Risk Assessment, where Natural Language Processing (NLP), financial analysis and Explainable Artificial Intelligence (XAI) intersect, has begun to attract the attention of researchers. However, the applications of LLMs in the field of credit risk analysis have not yet been systematically examined. The majority of studies in the literature superficially examine the applications of artificial intelligence in the financial sector or do not address the interpretability of LLMs in credit risk assessment in detail. Moreover, this particular research area requires a comprehensive taxonomy of LLM model types, data sources and explainable artificial intelligence (XAI) techniques.

\subsection{Objectives and Contributions}

This paper systematically reviews LLM-based approaches in credit risk assessment and presents a detailed taxonomy study. The main contributions of this paper are as follows:

\begin{itemize}

\item Published studies for LLM-based credit risk applications are systematically reviewed with the PRISMA methodology.

\item A detailed taxonomy is presented by examining the current research from four main aspects: model architecture, data modality, explainability mechanism and application area.

\item Trends, challenges and open research directions in the deployment of LLMs are analyzed.

\item The first systematic review and taxonomy focusing on LLM-based approaches in credit risk estimation is presented, providing a reference for researchers and financial institutions in the field.
 
\end{itemize}

\subsection{Paper Structure} 

\begin{figure}[ht]
	\centering
    \includegraphics[scale=0.5]{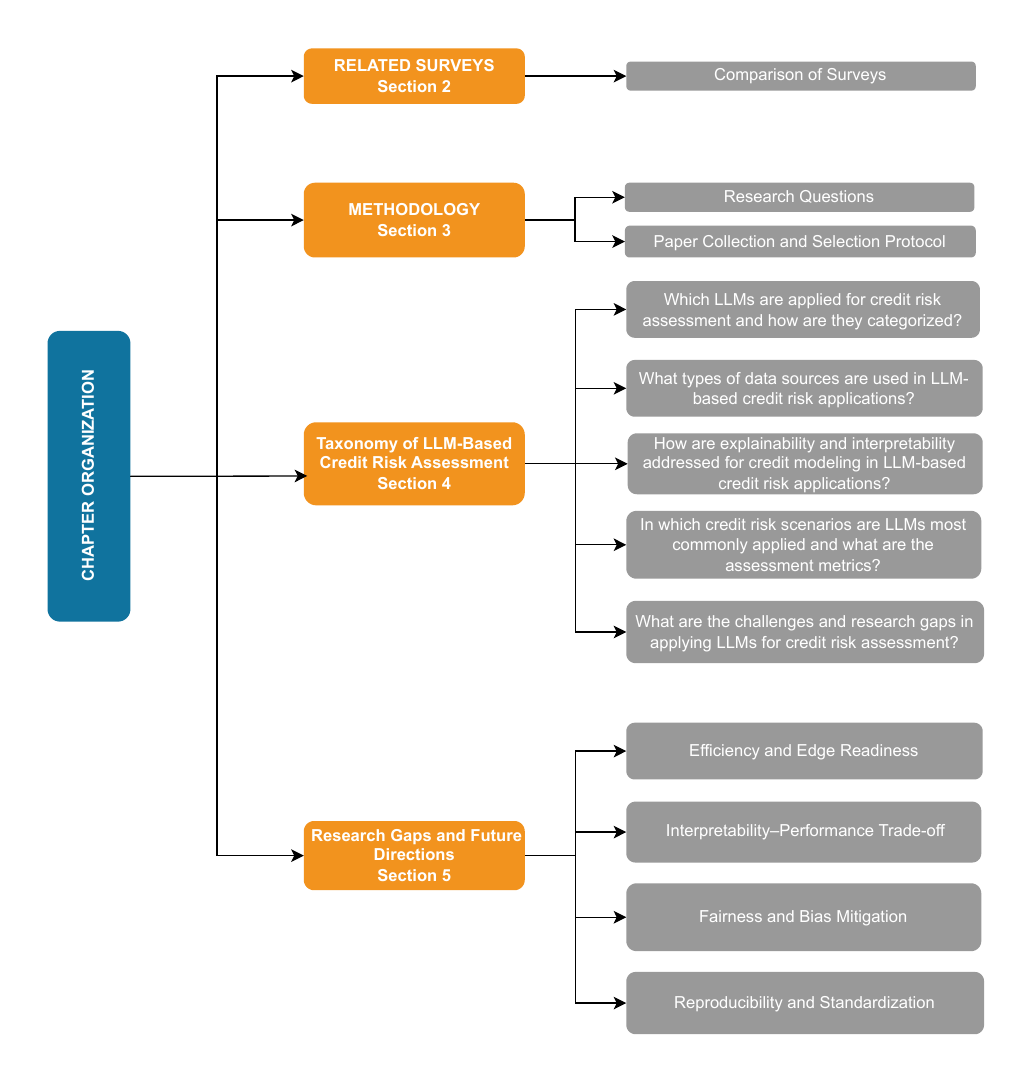}
	\caption{The Organization of the Paper}
	\label{fig:tax}
\end{figure}

Figure \ref{fig:tax} shows the organization of the paper. Section \ref{sec:2} examines the current survey studies in the field related finance and LLM , examining their focus and limitations. Then, a comparison analysis is performed by comparing this paper and the literature. Section \ref{sec:3} explains the research questions and article collection strategy. Section \ref{sec:4} provides a taxonomy by classifying LLM-oriented credit risk assessment systems in four main aspects: model architecture, data modality, interpretability mechanism and application area. Section \ref{sec:5} synthesizes the main emphases obtained throughout the paper and highlights the literature gaps. Section \ref{sec:6} concludes the paper by discussing the implications for researchers and finance practitioners in the field.

\section{Related Surveys} \label{sec:2}

Although there has been increasing interest in LLM and artificial intelligence (AI)-based financial studies in recent years, there is still no comprehensive classification in terms of focus, depth and relevance to credit risk, and discussions about interpretability, a key component in credit assessment scenarios, are lacking.

Staegemann et al. \cite{1}, in their literature review on generative AI applications in banking, highlight the potential of LLMs for risk reduction and customer experience. However, the article does not include techniques for credit risk modeling or interpretability, and there is no taxonomy of model types. In \cite{2}, models such as FinGPT and BloombergGPT are discussed, LLMs address financial inclusion and policy implications, and ethical concerns, but ignore credit risk applications. In Joshi et al. study's AI frameworks in credit risk and trading applications are examined, but does not include LLM architectures or datasets. \cite{3}. Chen et al. \cite{4}, in their study of LLMs in finance, law, and healthcare, also highlight the risks of using LLMs in high-risk sectors. However, the financial discussion is superficial, and credit scoring methods are not addressed. In \cite{5}, sentiment analysis techniques used in banking are reviewed, and the focus is on examining the impact on investor confidence. However, credit risk tasks are only superficially examined. In Nie et al. study's the  research on generative AI and LLMs focused on risk modeling and deployment architectures \cite{6}. The paper discusses areas such as data engineering, credit scoring modules, etc. but lacks a comparative taxonomy and interpretability mechanisms for LLM types. Zavitsanos et al. \cite{7} examines ML approaches in financial risk detection, examining features, data labels, evaluation techniques, and ML methods. Although this is a useful taxonomy, it is more focused on ML than LLM. In \cite{8}, the authors mainly emphasize traditional ML and statistical-based methods for financial forecasting, but do not address large language models or explainable AI methods. In Joshi et al. study's research on AI frameworks in credit risk and trading applications, but does not focus on LLM-specific taxonomy or performance benchmarking \cite{9}. Kong et al. \cite{10} presented a study that benchmarked LLM-based financial tasks for three different languages (Chinese, Japanese, English). However, the model types are not classified and also the explainability is not analyzed for credit risk domains. In Joshi et al. study's \cite{11}. provides detailed architectural review on LLM-based applications for credit scoring and macroeconomic simulations. However, the paper lacks empirical comparison and also lacks a classification for interpretability. Lee et al. \cite{12} review domain-specific financial applications based on LLM, but categorizes flow tasks and datasets, but does not focus on credit risk subdomains. In \cite{13} investigates AI-based methods in financial scenarios such as fraud detection and portfolio management, but does not provide a detailed review of model interpretability and LLM for credit scoring. In Karami and Igbokwe study examines AI-based risk assessment and limitations of traditional credit methods, but does not provide a classification of LLM \cite{14}. Alonso et al. \cite{15} provides an overview of LLMs in areas such as asset management and risk reporting, but is insufficient for credit risk. In \cite{16}, LLMs are studied for financial applications such as credit analysis and fraud detection. However, they are very superficial in interpretability and credit-specific modeling. Lakkaraju et al. \cite{17} evaluates LLM performance through a fairness lens, focuses on bias reduction strategies such as improving user trust, but credit risk assessments and explainability are very superficial. In \cite{18}, LLM-based financial sentiment analysis with multiple datasets is not considered credit risk or XAI. Omoseebi et al. \cite{19} investigates the financial security side of LLMs such as fraud detection, but does not perform detailed credit risk modeling and taxonomy studies. In \cite{20}, democratization of financial datasets and open data models of LLMs are discussed, but no empirical comparison is included for credit or risk analytics. Krause et al. \cite{23} present a survey discussing the sustainability of LLM models such as ChatGPT. The survey is mostly on ethical and governance concerns and is superficial in terms of credit scoring, model breakdowns, and datasets. In \cite{25}, the robustness of LLMs in financial tasks is assessed and benchmarks are made on summarization and event detection. However, the study does not include credit-related tasks and explainability is weak.

\begin{table}[h]
\caption{Comparison of Related Survey Studies with This Work}
\label{tab:survey_comparison}
\resizebox{\textwidth}{!}{%
\begin{tabular}{@{}cccccc@{}}
\toprule
\textbf{Work}      & \textbf{Main Aim}               & \textbf{Credit Risk} & \textbf{LLM-Specific} & \textbf{XAI} & \textbf{Taxonomy} \\ \midrule
\cite{1}                  & GenAI in Banking                & \ding{55}                  & \ding{51}                   & \ding{55}           & \ding{55}                \\
\cite{2}                  & LLMs for Financial Regulation   & \ding{55}                   & \ding{51}                   & \ding{55}           & \ding{55}                \\
\cite{3}                  & Agentic GenAI for Risk Mgmt     & \ding{51}                  & \ding{51}                   & \ding{55}           & \ding{55}                \\
\cite{4}                  & LLMs in Finance/Healthcare/Law  & \ding{55}                   & \ding{51}                   & \ding{55}           & \ding{55}                \\
\cite{5}                  & Sentiment in Banking Headlines  & \ding{55}                   & \ding{51}                   & \ding{55}           & \ding{55}                \\
\cite{6}                  & LLMs in Financial Applications  & \ding{55}                   &\ding{55}                   & \ding{55}           & \ding{55}                \\
\cite{7}                  & ML for Financial Risk Reports   & \ding{51}                  & \ding{55}                    & \ding{55}           & \ding{51}               \\
\cite{8}                  & AI in Modern Banking            & \ding{51}                  & \ding{51}                   & \ding{55}           & \ding{55}                \\
\cite{9}                  & GenAI Agents in Finance         & \ding{51}                  & \ding{51}                   & \ding{55}           & \ding{55}                \\
\cite{10}                 & LLMs in Investment Mgmt         & \ding{55}                   & \ding{51}                   & \ding{55}           & \ding{55}                \\
\cite{11}                 & GenAI for Financial Risk        & \ding{51}                  & \ding{51}                   & \ding{55}           & \ding{55}                \\
\cite{12}                 & Survey of FinLLMs               & \ding{51}                  & \ding{51}                   & \ding{55}           & \ding{51}               \\
\cite{13}                 & Explainable AI in Finance       & \ding{51}                  & \ding{55}                    & \ding{51}          & \ding{55}                \\
\cite{14}                 & Big Data in Credit Risk         & \ding{51}                  & \ding{55}                    & \ding{55}           & \ding{51}               \\
\cite{15}                 & LLMs for Financial Reasoning    & \ding{55}                   & \ding{51}                   & \ding{55}           & \ding{55}                \\
\cite{16}                 & LLMs in Financial AI            & \ding{55}                   & \ding{51}                   & \ding{55}           & \ding{51}               \\
\cite{17}                 & LLMs as Finance Advisors        & \ding{55}                   & \ding{51}                   & \ding{55}           & \ding{55}                \\
\cite{18}                 & LLMs for Sentiment Analysis     & \ding{55}                   & \ding{51}                   & \ding{55}           & \ding{55}                \\
\cite{19}                 & LLMs for Financial Security     & \ding{55}                   & \ding{51}                   & \ding{55}           & \ding{55}                \\
\cite{20}                 & Consistency of LLMs             & \ding{55}                   & \ding{51}                   & \ding{55}           & \ding{55}                \\
\cite{23}                 & LLMs in Finance (ChatGPT/Bard)  & \ding{55}                   & \ding{51}                   & \ding{51}          & \ding{55}                \\
\cite{25}                 & LLM Strategy in FIs             & \ding{55}                   & \ding{51}                   & \ding{55}           & \ding{55}                \\
\textbf{This Work} & Interpretable LLMs for Credit Risk & \textbf{\ding{51}}         & \textbf{\ding{51}}          & \textbf{\ding{51}} & \textbf{\ding{51}}      \\ \bottomrule
\end{tabular}%
}
\end{table}

\textbf{Comparison of Surveys}: Table~\ref{tab:survey_comparison} provides a comparison of the surveys examined in this section. While interest in the financial domain applications of LLMs is increasing day by day, existing studies focus on broad themes such as generative AI \cite{1}, inclusiveness frameworks \cite{2} and financial NLP \cite{12, 16}. Only a limited number of studies directly focus on credit risk tasks, and none of them provide a taxonomy by examining LLM architectures, data formats, interpretability techniques and domain-specific applications.

Furthermore, most studies do not provide interpretability for high-risk decisions such as credit scoring, and the studies that do provide interpretability are very superficial \cite{15, 20}. Some of the surveys include studies such as sentiment analysis or fraud detection \cite{18, 5}, but they do not address the concept of credit risk modeling. Some of the reviewed studies focus only on traditional machine learning (ML) \cite{7, 14} and only propose conceptual frameworks without model comparison \cite{3, 11}.

Although some of the studies discuss domain-specific LLMs (such as FinBERT, FinGPT, and InvestLM), they do not systematically model architecture or risk. None of the existing studies provide a classification of LLM applications that addresses both modeling and regulatory requirements (e.g., explainability, data provenance, or fairness). \textbf{This study is the first to fill this gap based on four main pillars: model architecture, data modality, interpretability mechanism, and application domain.}

\section{Methodology} \label{sec:3}

This section describes the methodology used in the systematic review of LLM-based credit risk assessment.

\subsection{Paper Collection and Selection Protocol}

For the paper collection, refereed journals, high-ranking conferences, book chapters covering the years 2020-2025 were collected by scanning libraries such as IEEE Xplore, Elsevier, ACM Digital Library, SpringerLink, Scopus and arXiv. The following keywords were used while scanning the paper:

\begin{enumerate}
  \item \verb|[(LLM) || (LargeLanguageModel)] \& [(CreditScoring) || (CreditRiskAssessment)]|
  \item \verb|[(LLM) || (Transformer)] \& [(Explainability) || (XAI)] \& [(FinancialNLP)]|
  \item \verb|[(CreditRisk)] & [(MultimodalData) || (UnstructuredText) || (BehavioralData)]|
  \item \verb|[(LLM)] & [(GPT4 || FinBERT)] \& [(Evaluation || Benchmark)]|
  \item \verb|[(Fairness) || (Hallucination) || (Reproducibility)] \& [(LLM) || (CreditRisk)]|
\end{enumerate}

\begin{figure}[h]
    \centering
    \includegraphics[width=0.85\linewidth]{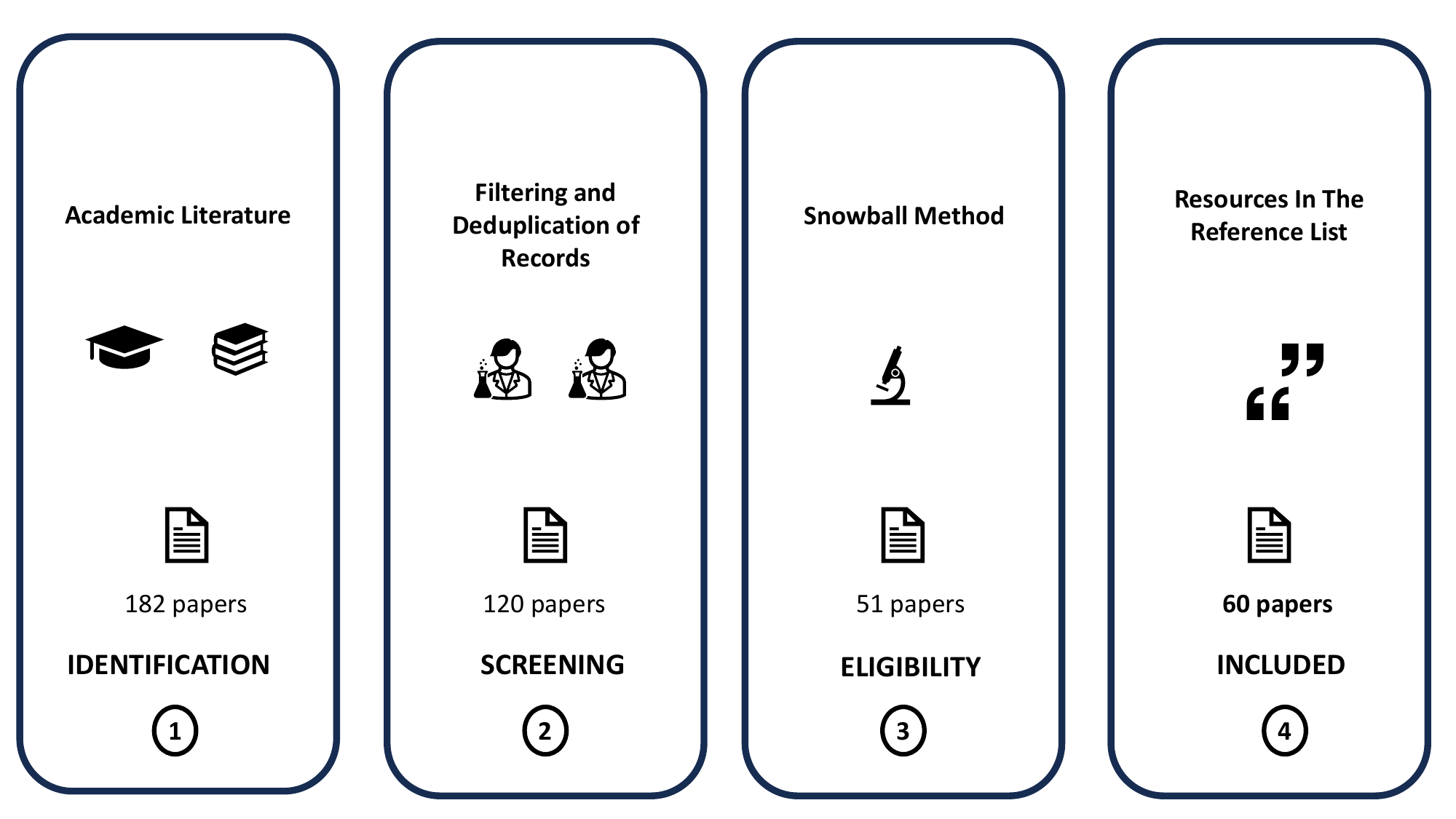}
    \caption{PRISMA flow diagram for the selection of studies on LLM-based credit risk assessment.}
    \label{fig:prisma}
\end{figure}

Figure \ref{fig:prisma} summarizes the strategy followed for paper collection in this survey. Following the PRISMA guidelines, 182 papers were first collected as the initial body as a result of the relevant keys. After removing duplicates and papers that were not related to the research area, 120 papers were obtained. With the joint efforts of both authors in this survey (according to the inclusion criteria), 51 articles were obtained for the final analysis and this number was increased to 60 papers with the snowball method. A classification of all these papers is shown in Table \ref{tab:overview_papers1}.

\begin{table*}[h]
\centering
\caption{Overview of 60 Selected Studies}
\label{tab:overview_papers1}
\scriptsize
\resizebox{\textwidth}{!}{%
\begin{tabular}{|c|c|c|c|}
\hline
\textbf{No} & \textbf{Paper}                       & \textbf{Venue/Platform}                                                                        & \textbf{Year}             \\ \hline
1           & Sanz-Guerrero et al., \cite{1a}          & Inteligencia Artificial (IBERAMIA)                                                             & 2025                      \\ \hline
2           & Dogra et al., \cite{2a}                  & MDPI Systems                                                                                   & 2022                      \\ \hline
3           & Dolphin et al., \cite{3a}                & arXiv / Polygon.io                                                                             & 2024                      \\ \hline
4           & Cai, \cite{4a}                           & IEEE ICEDCS                                                                                    & 2024                      \\ \hline
5           & Govindaraj et al., \cite{5a}             & World Journal of Advanced Research Reviews                                                     & 2023                      \\ \hline
6           & Xie et al., \cite{6a}                    & arXiv (Preprint under review)                                                                  & 2023                      \\ \hline
7           & Babaei \& Giudici, \cite{7a}             & Machine Learning with Applications (Elsevier)                                                  & 2024                      \\ \hline
8           & Loukas et al., \cite{8a}                 & ACM ICAIF '23 (International Conference on AI in Finance)                                      & 2023                      \\ \hline
9           & Liu et al., \cite{9a}                    & ACM DEBAI 2024 (Digital Economy, Blockchain \& AI)                                             & 2024                      \\ \hline
10          & Mehedi Hasan et al., \cite{10a}          & International Journal of Computer Science \& Info Systems                                      & 2024                      \\ \hline
11          & Charlie Luca, \cite{11a}                 & ResearchGate (Preprint)                                                                        & 2024                      \\ \hline
12          & Pau Rodriguez Inserte et al., \cite{12a} & arXiv                                                                                          & 2024                      \\ \hline
13          & Linyi Yang et al., \cite{13a}            & arXiv                                                                                          & 2020                      \\ \hline
14          & Jiarui Rao \& Qian Zhang, \cite{14a}     & International Journal of Multidisciplinary Research and Growth Evaluation                      & 2025                      \\ \hline
15          & Sungwook Yoon, \cite{15a}                & International Journal of Advanced Smart Convergence                                            & 2023                      \\ \hline
16          & Duanyu Feng et al., \cite{16a}           & ACM (Conference acronym ’XX)                                                                   & 2024 \\ \hline
17          & Ayomide Joel et al., \cite{17a}          & ResearchGate / Unpublished                                                                     & 2023                      \\ \hline
18          & Jaskaran Singh Walia et al., \cite{18a}  & arXiv (arXiv:2502.17011v1)                                                                     & 2025                      \\ \hline
19          & Xue Wen Tan and Stanley Kok, \cite{19a}  & ICIS (AIS Electronic Library)                                                                  & 2023                      \\ \hline
20          & Khaoula Idbenjra et al., \cite{20a}      & Elsevier – Decision Support Systems                                                            & 2024                      \\ \hline
21          & Zhang et al., \cite{21a}                 & arXiv                                                                                          & 2023                      \\ \hline
22          & Malaysha et al., \cite{22a}              & arXiv                                                                                          & 2024                      \\ \hline
23          & Wu et al., \cite{23a}                    & CRC Working Papers                                                                             & 2024                      \\ \hline
24          & Sideras et al., \cite{24a}               & ACM (ICAIF ’24)                                                                                & 2024                      \\ \hline
25          & Fatemi et al., \cite{25a}                & arXiv                                                                                          & 2024                      \\ \hline
26          & Kalluri et al, \cite{26a}                & IJNRD                                                                                          & 2024                      \\ \hline
27          & Lin et al, \cite{27a}                    & ACM ICAIF                                                                                      & 2024                      \\ \hline
28          & Lei et al, \cite{28a}                    & arXiv                                                                                          & 2025                      \\ \hline
29          & Lakkaraju et al, \cite{29a}              & ACM ICAIF                                                                                      & 2023                      \\ \hline
30          & Liu et al, \cite{30a}                    & NeurIPS Workshop                                                                               & 2023                      \\ \hline
31          & Lopez-Lira et al., \cite{31a}            & arXiv                                                                                          & 2025                      \\ \hline
32          & Xie et al., \cite{32a}                   & NeurIPS (Datasets \& Benchmarks Track)                                                         & 2023                      \\ \hline
33          & Huang et al., \cite{33a}                 & MDPI Applied Sciences                                                                          & 2025                      \\ \hline
34          & Moraes et al., \cite{34a}                & WebMedia (Brazilian Symposium on Multimedia and the Web)                                       & 2024                      \\ \hline
35          & Huang et al., \cite{35a}                 & arXiv                                                                                          & 2025                      \\ \hline
36          & Busireddy et al., \cite{36a}             & Int. Jr. of Hum Comp. \& Int.                                                                  & 2025                      \\ \hline
37          & Babaei \& Giudici, \cite{37a}           & Machine Learning with Applications                                                             & 2024                      \\ \hline
38          & Wang et al., \cite{38a}                 & IEEE Transactions on Engineering Management                                                    & 2025                      \\ \hline
39          & Lin et al., \cite{39a}                   & Journal of Data, Information and Management                                                    & 2025                      \\ \hline
40          & Hartomo et al., \cite{40a}               & IEEE Access                                                                                    & 2025                      \\ \hline
41          & Yan et al., \cite{41a}                   & IEEE RAAI (Robotics, Automation, and AI Conference)                                            & 2024                      \\ \hline
42          & Gupta et al., \cite{42a}                 & JPMorgan Chase (Internal ML Research)                                                          & 2024                      \\ \hline
43          & Suresh et al., \cite{43a}                & Educational Administration: Theory and Practice (Kuey)                                         & 2024                      \\ \hline
44          & Ni et al., \cite{44a}                    & IEEE DOCS (Data-driven Optimization of Complex Systems)                                        & 2024                      \\ \hline
45          & Bond et al., \cite{45a}                  & Working Paper / Preprint (University of Queensland)                                            & 2024                      \\ \hline
46          & Sun et al., \cite{46a}                   & IEEE Access                                                                                    & 2024                      \\ \hline
47          & Sabuncuoglu et al., \cite{47a}           & IEEE Symposium on Computational Intelligence for Financial Engineering and Economics (CIFEr)   & 2025                      \\ \hline
48          & Xie et al., \cite{48a}                   & NeurIPS (Track on Datasets and Benchmarks)                                                     & 2024                      \\ \hline
49          & Liu et al., \cite{49a}                  & IEEE 4th International Conference on Computer Communication and Artificial Intelligence (CCAI) & 2024                      \\ \hline
50          & Liu et al., \cite{50a}                   & \textit{Finance Research Letters (Elsevier)}                                                   & 2025                      \\ \hline
51          & Chanda and Prabhu, \cite{51a}            & \textit{IEEE ICICCS Conference Proceedings}                                                    & 2023                      \\ \hline
52          & Papasotiriou et al., \cite{52a}          & \textit{ACM ICAIF ’24 (AI in Finance Conference)}                                              & 2024                      \\ \hline
53          & Rizinski et al., \cite{53a}              & \textit{IEEE Access}                                                                           & 2024                      \\ \hline
54          & Chafekar et al., \cite{54a}              & \textit{Unspecified (likely arXiv or workshop preprint)}                                       & 2024                      \\ \hline
55          & Li et al., \cite{55a}                    & SSRN                                                                                           & 2024                      \\ \hline
56          & Kim et al., \cite{56a}                   & ACM ICAIF                                                                                      & 2023                      \\ \hline
57          & Sanz-Guerrero et al., \cite{57a}         & SSRN                                                                                           & 2024                      \\ \hline
58          & Guo et al., \cite{58a}                   & arXiv                                                                                          & 2023                      \\ \hline
59          & Fallahgoul, \cite{59a}                   & SSRN (Preprint)                                                                                & 2025                      \\ \hline
60          & Zhang et al., \cite{60a}                 & ACM ICAIF                                                                                      & 2023                      \\ \hline
\end{tabular}%
}
\end{table*}

\subsection{Research Questions}

In this paper, the systematic review is structured around five research questions. Table \ref{tab:research_questions} shows the research questions, motivation, and relevant section information for this paper.

\begin{table}[h]
\caption{The Research Questions of the Survey}
\label{tab:research_questions}
\resizebox{\textwidth}{!}{%
\begin{tabular}{ccc}
\hline
\textbf{Research Question} & \textbf{Motivation} & \textbf{Section} \\ \hline
\begin{tabular}[c]{@{}c@{}}Which LLMs are applied for \\ credit risk assessment and how are \\ they categorized?\end{tabular} &
\begin{tabular}[c]{@{}c@{}}The purpose is to understand LLM architectures \\ used for credit risk applications \\ and identify performance bottlenecks \\ and suitability for domain-specific tasks.\end{tabular} &
4.1 \\
\begin{tabular}[c]{@{}c@{}}What types of data sources \\ are used in LLM-based credit risk \\ applications?\end{tabular} &
\begin{tabular}[c]{@{}c@{}}The purpose is to evaluate how \\ comprehensive mapping of data \\ sources provides model generalization \\ and fairness.\end{tabular} &
4.2 \\
\begin{tabular}[c]{@{}c@{}}How are explainability and \\ interpretability addressed for \\ credit modeling in LLM-based credit risk \\ applications?\end{tabular} &
\begin{tabular}[c]{@{}c@{}}The purpose is to evaluate \\ how model transparency and accountability are \\ addressed in existing studies.\end{tabular} &
4.3 \\
\begin{tabular}[c]{@{}c@{}}In which credit risk scenarios \\ are LLMs most commonly applied and \\ what are the assessment metrics?\end{tabular} &
\begin{tabular}[c]{@{}c@{}}The purpose is to identify common use \\ cases of existing studies and thus provide insight into \\ practical relevance, dataset diversity, and reproducibility.\end{tabular} &
4.4 \\
\begin{tabular}[c]{@{}c@{}}What are the challenges and \\ research gaps in applying LLMs for \\ credit risk assessment?\end{tabular} &
\begin{tabular}[c]{@{}c@{}}The purpose is to guide future research \\ by highlighting challenges and open questions.\end{tabular} &
5 \\ \hline
\end{tabular}%
}
\end{table}

\section{Taxonomy of LLM-Based Credit Risk Assessment}\label{sec:4}

The papers collected in this section are classified around four main categories: model architectures, data formats, interpretability mechanisms and application areas, and answers are sought to RQ1, RQ2, RQ3, and RQ4.   

\subsection{Model Architectures (RQ1)}

\begin{figure}[h]
    \centering
    \includegraphics[width=0.5\linewidth]{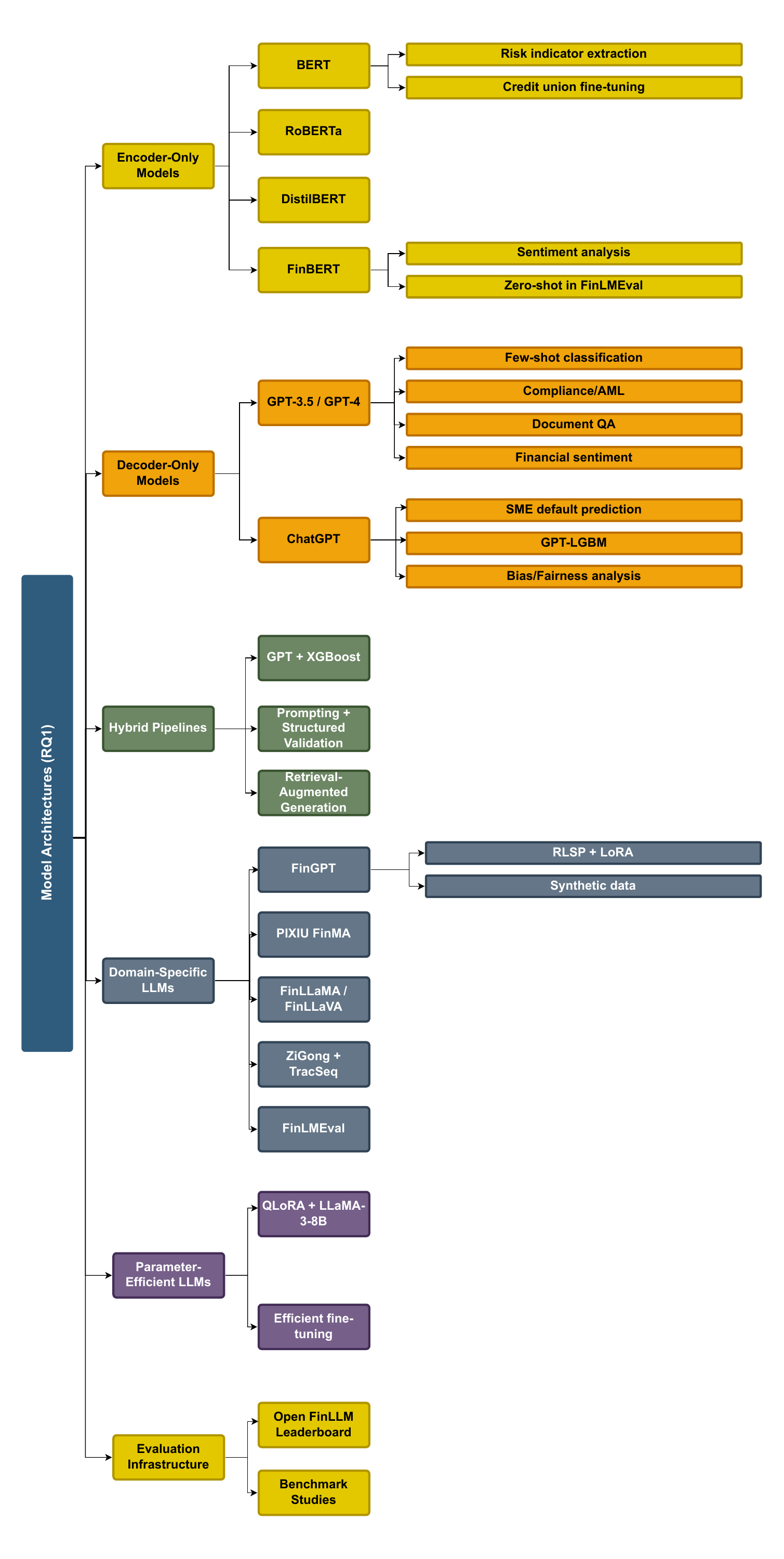}
    \caption{Taxonomy of Transformer-Based Model Architectures for LLM-Driven Credit Risk Assessment.}
    \label{fig:taxonomy_rq1_models}
\end{figure}

Recent research on LLM-based credit risk has spanned a variety of domains, from transformer-based architectures to hybrid pipelines and domain-specific fine-tuning to efficient learning. This subsection surveys model architectures for credit risk prediction found in the literature to answer RQ1. Figure \ref{fig:taxonomy_rq1_models} shows a taxonomy of these models.

\begin{itemize}

    \item \textbf{Encoder-Only Architectures}: Encoder-only models (RoBERTa, DistilBERT, FinBERT) are widely used in textual analysis-based financial classification \cite{1a}. The BERT model derives credit risk indicators and is used to integrate them with tree-based learners \cite{1a}. FinBERT is reported to perform well in financial sentiment prediction \cite{36a} and fine-tuning in encoder-based benchmarking \cite{58a}. Luca studies the BERT model for credit union use in member transactions \cite{11a}. It is also reported that BERT performs robust sentiment extraction in multilingual complex environments \cite{56a}.

    \item \textbf{Decoder-Only Architectures}: Decoding models (GPT-3.5, GPT-4 and ChatGPT) are widely used in generative tasks. Loukas et al. \cite{8a} showed the performance of these models (GPT-3.5 and GPT-4) in financial intent classification with a small number of examples (8-20). Similarly, Babaei and Giudici \cite{7a,37a} report the success of GPT in low-data credit scoring scenarios.

    ChatGPT has been tested by applying it to multidimensional prediction models (financial ratios, social media sentiment) \cite{39a}. Its psychological feature extraction capability is investigated and demonstrated in the GPT-LGBM framework \cite{55a}. However, Gupta et al. \cite{42a} mention limitations such as following instructions in complex documents for long-context environments in their study using GPT-4-Turbo.

    \item \textbf{Hybrid and Augmented Pipelines}: Some frameworks are seen to combine transform-based language modeling with traditional ML models/retrieval mechanisms. In \cite{1a}, BERT and GPT are seen to be used together with XGBoost. Another work introduces a structured request pipeline in financial disclosures \cite{3a}. RAG architectures have gained attention for low-context financial data tasks (news sentiment classification) \cite{8a,60a}. All these approaches aim to improve the response base by integrating external information sources.

    \item \textbf{Domain-Specific Financial LLMs (FinLLMs)}: Models such as FinGPT, FinMA, ZiGong and FinLLaMA have been built to address domain variations in financial texts. FinGPT is a low-cost framework based on LoRA and Stock Price Reinforcement Learning \cite{30a}. Another study initiates the Open FinLLM Leaderboard when comparing financial LLMs and contributes to this field \cite{27a}. FinMA framework outperforms GPT-4 on credit-related tasks \cite{32a}. ZiGong \cite{28a} aims to reduce hallucinations with TracSeq using instruction tuning and temporal pruning methods. FinLMEval emphasizes that the fine-tuning ability of encoder models is superior to the zero-shot power of decoder models in data-scarce environments \cite{58a}. The Open-FinLLMs package supports textual, tabular and visual financial inputs for representing advances in multimodal credit modelling \cite{35a}.

    \item \textbf{Parameter-Efficient LLMs and Training Techniques}: To reduce the computational cost of large models, recent studies have emphasized parameter-efficient methods. Kalluri \cite{26a} proposed a study that improves the interaction accuracy and compatibility rates with a scalable solution. In another study, Ni et al. \cite{44a} proposes QLoRA-based tuning with LLaMA-3-8B-Instruct-4bit compact models. It is emphasized that the results are better than GPT-4 for gain data.

    \item \textbf{Benchmarking and Evaluation Tools}: Many recent studies also aim to standardize LLM evaluation in finance. Examples include FinLLM Leaderboard \cite{27a}, which allows for repeatable testing of financial benchmarks, FinLMEval \cite{58a}, which provides analysis of encoder and decoder performance for fine-tuning and zero-shotting. In another study, Lakkaraju et al. \cite{29a} emphasize the concept of fairness for bias models for ChatGPT and Bard outputs.
    
\end{itemize}

\subsection{Data Modalities (RQ2)}

\begin{figure}[h]
    \centering
    \includegraphics[width=0.5\linewidth]{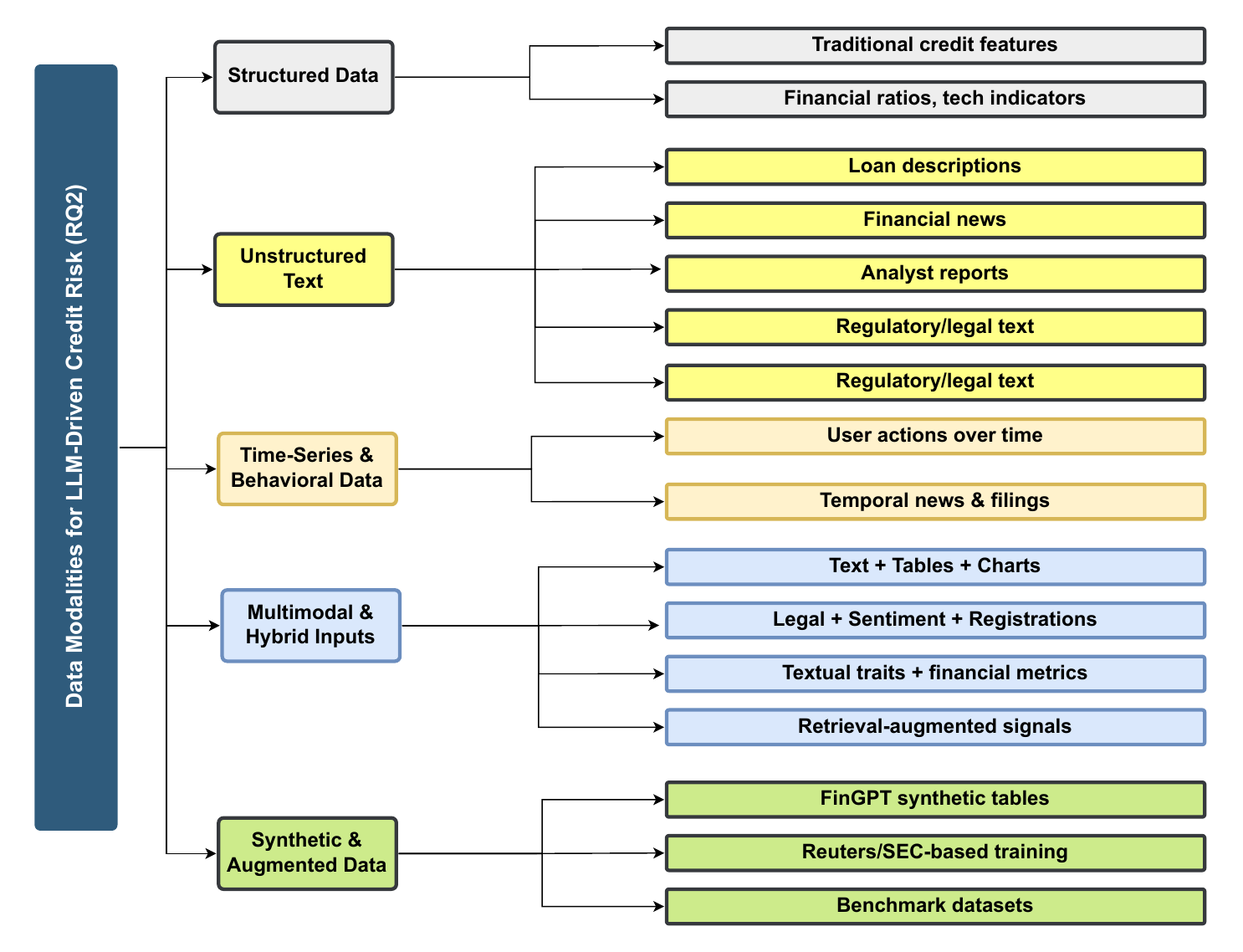}
    \caption{Taxonomy of Data Modalities Utilized in LLM-Based Credit Risk Systems.}
    \label{fig:taxonomy_rq2_data}
\end{figure}

Recent developments have also witnessed major improvements in the data processed for LLM-focused credit risk assessment. While traditional credit modeling was done with structured tabular data, various data modalities have now emerged, such as unstructured financial text, time series behaviors, multimodal data pipelines, and integration of synthetic data. Figure \ref{fig:taxonomy_rq2_data} shows the taxonomy of data modalities.

\begin{itemize}
    
\item \textbf{Structured Data}: Structured data such as income level and defaults are frequently used in credit scoring studies. GPT-LGBM aims to improve classification performance using this data and ChatGPT \cite{55a}. Similarly, PIXIU's FLARE benchmark and Open-FinLLMs pre-training body use structured data such as technical indicators and historical prices \cite{32a,35a}.

\item \textbf{Unstructured Financial Text}: Recent studies have reported that unstructured data such as credit disclosures and regulatory documents are important in capturing hidden credit indicators. Sanz-Guerrero et al. \cite{1a}, Dogra et al. \cite{2a} built enhanced models with free-text disclosures and news sentiment. In other studies, Wu et al. \cite{23a} and Sideras et al. \cite{24a} examined the impact of manager and auditor comments on default accuracy.

While ChatGPT can be used to calculate market sentiment scores using long-form news \cite{45a}, it has been shown to be used effectively with GPT/BERT in the compliance profiling of AML-related documents \cite{41a}. Relatively low-resource models (such as GPT-3.5) have been shown to outperform traditional models in extracting sentiment from analyst reports \cite{56a}

\item \textbf{Time-Series and Behavioral Data}: Another data type that is starting to be integrated into LLM pipelines is time series. Lei et al. \cite{28a} filter out low quality sequences in the ZiGong model with a pruning strategy and show the importance of user activity data (time series). Another case study using this data structure is FinGPT, where real-time time series data is used \cite{30a}.

\item \textbf{Multimodal and Hybrid Inputs}: Textual and behavioral features can be combined for more comprehensive credit modeling. Huang et al. \cite{33a} achieved high prediction accuracy using claim-based LLMs and records of 38 real-world SMEs. Another study integrates psychological features extracted from texts with financial variables in GPT-LGBM \cite{55a}. Another hybrid input example is the study by Zhang et al. \cite{60a} where LLMs are used to enrich external information by combining context-deficient financial text.

\item \textbf{Synthetic and Augmented Data}: Factors affecting model performance such as data incompleteness and class imbalances can be addressed using synthetic data-generated LLM models. FinGPT is used to simulate credit records \cite{4a}, another example is domain-adaptive LLM models trained on SEC filings and Reuters headlines \cite{12a}. Similarly, Feng et al. \cite{16a} proposed benchmark models for fraud and bankruptcy-related datasets, and the results show success in cross-modality learning.

\end{itemize}

\subsection{Interpretability Mechanisms (RQ3)}

\begin{figure}[h]
    \centering
    \includegraphics[width=0.5\linewidth]{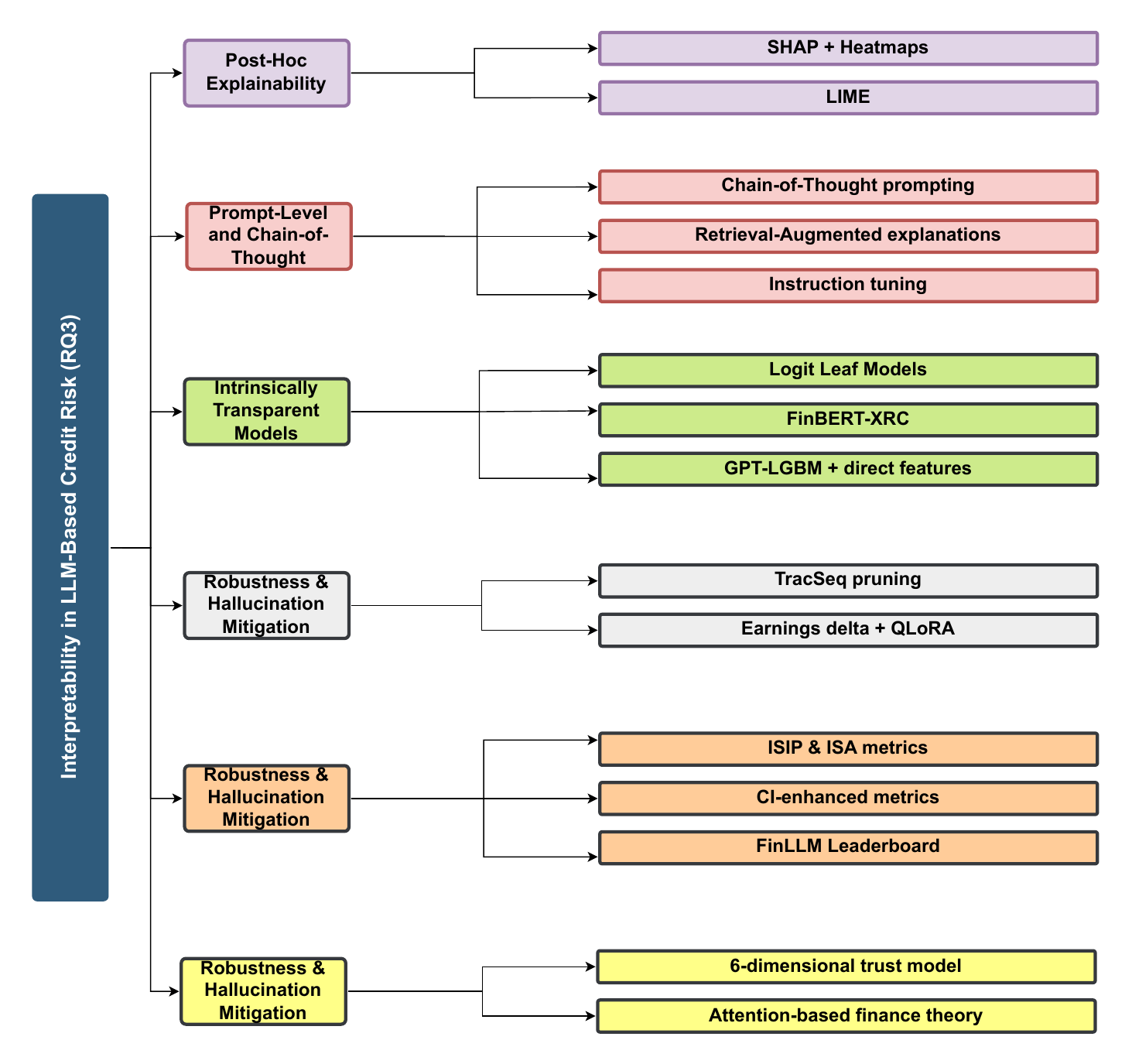}
    \caption{Taxonomy of Interpretability Mechanisms for LLM-Based Credit Models.}
    \label{fig:taxonomy_rq3_interpretability}
\end{figure}

Interpretability is an important concept in LLM-based credit risk assessment due to its fairness and reliability. Recent studies aim to increase transparency with methods such as post-hoc methods, internal model designs, instruction setting and model checking strategies. Figure \ref{fig:taxonomy_rq3_interpretability} shows the taxonomy of interpretability mechanisms.

\begin{itemize}
    
\item \textbf{Post-Hoc Explainability}: The most commonly used post-hoc methods in this method are SHAP and LIME. Govindaraj et al. \cite{5a} proposed a study to visualize local global feature contributions by applying SHAP and attention heatmaps. Liu et al. \cite{9a} conducted a study combining SHAP and LIME outputs with ChatGPT to make them human interpretable.In \cite{23a}, LIME's efficiency in credit assessments is examined, while in \cite{40a} SHAP is used with TabTransformer for class skew compensation and interpretability.

\item \textbf{Chain-of-Thought and Prompt-
Level Explanations}: Recent work aims to enable models to generate explanations directly in the inference pipeline. Dolphin et al. \cite{3a} use chain-of-thought prompts to provide interpretability in sentiment classification scenarios, while in \cite{8a} they use backoff to improve prediction fixation and tractability. Fatemi et al. \cite{25a} show that interpretability can be increased by improving the model’s internal reasoning through instruction tuning.

\item \textbf{Intrinsically Interpretable 
Model Designs}: Some studies propose transparent architectures to reduce the dependency on external explainability tools. One of these models is Logit Leaf, which integrates LLM outputs and segmentation trees, as proposed by Idbenjra et al. \cite{20a}.  Another study introduces FinBERT-XRC, which can interpret risks at the word and sentence level  \cite{19a} . Li et al. \cite{55a} reported that the personality traits extracted by GPT-LGBM are directly interpretable (without post-hoc XAI).

\item \textbf{Robustness and Hallucination 
Mitigation}: One of the factors that directly affects interpretability is model robustness. To increase model robustness and address hallucinations, in \cite{28a}, the authors propose a model called TracSeq. In another study, the authors aim to improve interpretability by combining gain deltas and the QLoRA model.

\item \textbf{Fairness, Auditing, and 
Reproducibility}: Other concepts that are directly related to interpretability are fairness and reproducibility. In \cite{29a} ISIP and ISA metrics are used to check model fairness. Gupta et al. \cite{42a} emphasize the importance of F1 score and confidence intervals (CI) metrics for explainability checks. Additionally, Lin et al. \cite{27a} advocate benchmarking with the FinLLM Leaderboard to standardize interpretability assessment.

\item \textbf{Ethical, Regulatory, and 
Theoretical Dimensions}: For reliable interpretability, ethics should also be taken into account. Yan et al. \cite{41a} highlights six basic dimensions of compliance to evaluate LLMs in critical applications: correctness, fairness, privacy, robustness, security, and ethics. Another paper \cite{59a} discusses how to extend the Markowitz framework to provide attention-based interpretability and optimization.

\end{itemize}

\subsection{Application Domains (RQ4)}

\begin{figure}[h]
    \centering
    \includegraphics[width=0.5\linewidth]{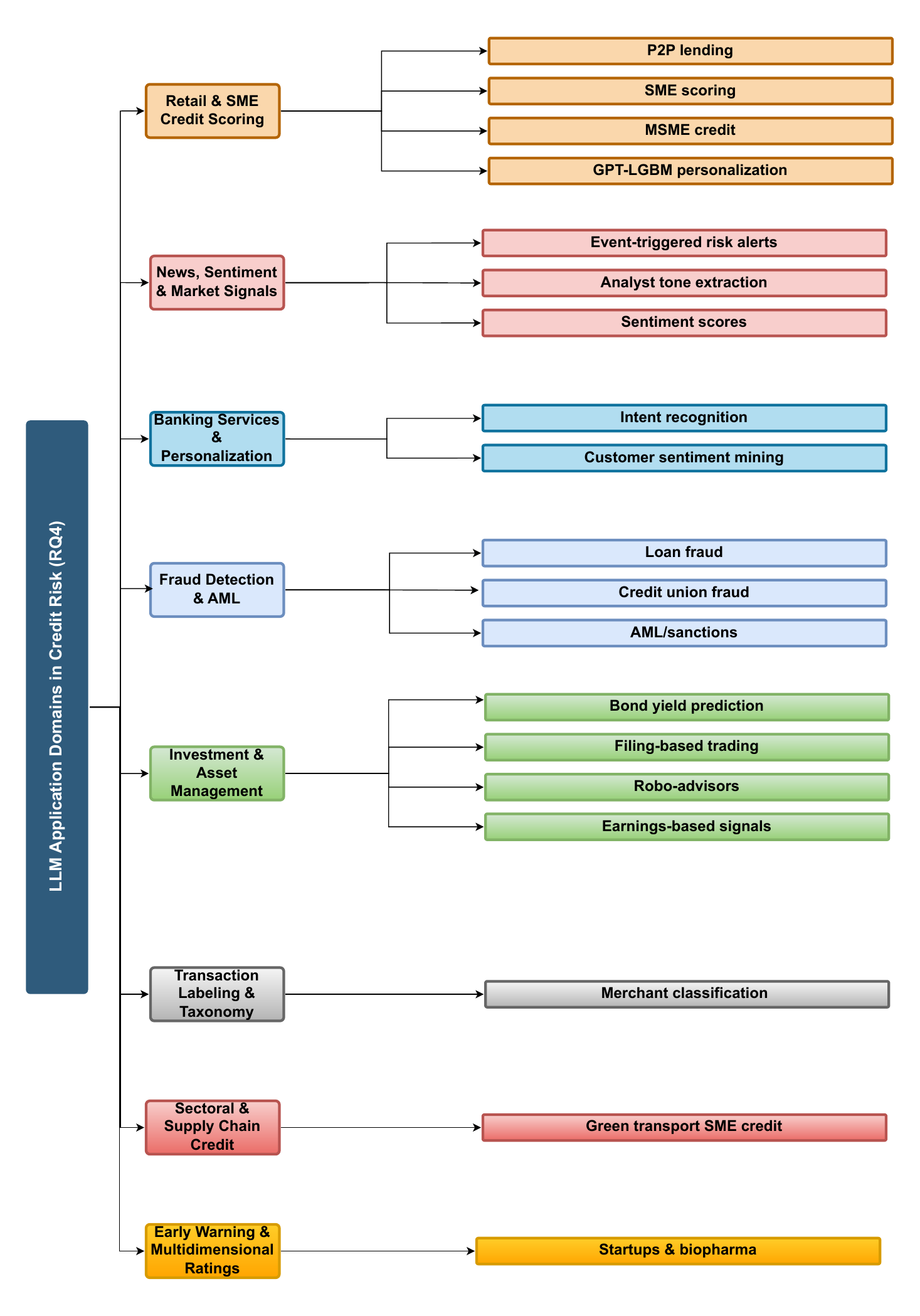}
    \caption{Taxonomy of Application Domains for LLMs in Credit Risk Assessment.}
    \label{fig:taxonomy_rq4_applications}
\end{figure}

LLMs have recently been applied in a wide range of areas, including traditional credit scoring, fraud detection, sentiment prediction, robo-advisory, etc. Figure \ref{fig:taxonomy_rq4_applications} shows the Taxonomy of Application Domains for LLMs in Credit Risk Assessment.

\begin{itemize}
    
\item  \textbf{Retail and SME Credit Scoring}: When the literature is examined, it is seen that LLMs are used in various financial application areas such as peer-to-peer (P2P) lending and SME financing decisions. In their studies, Sanz-Guerrero et al. \cite{1a} and Babaei \& Giudici \cite{7a} use credit disclosures for borrower creditworthiness classification. In \cite{15a}, SME technology loans are evaluated with a GPT-based credit assessment mechanism. In another study, \cite{33a}, claims-based assessments are applied to non-financial data for SME default prediction. In \cite{40a}, the SHAP-TabTransformer model is applied to SME credit classification. Li et al. \cite{55a} combine structured financial data with extracted personality traits for credit assessment in their proposed GPT-LGBM framework.

\item  \textbf{Financial News, Sentiment, and Market Signals}: LLMs performed well in extracting market sentiment and event triggers even on unstructured financial data. Dolphin et al. \cite{3a} and Dogra et al. \cite{2a} used LLMs to predict risk from headlines with event triggers. In \cite{45a} daily S\&P 500 sentiment scores were generated using GPT-3.5, while in another study \cite{56a} analyst tone was measured. Seshakagari et al \cite{36a} used GPT-4 to classify financial sentiment in a changing macroeconomic environment.

\item  \textbf{Customer Service and Banking Operations}: Some studies in the literature focus on intent recognition and service personalization with LLMs. Loukas et al. \cite{8a} performed service request classification with the Banking77 dataset, while in \cite{43a} customer sentiment analysis was used to support banking service personalization and credit decisions.

\item  \textbf{Fraud Detection and Anti-Money Laundering (AML)}: FinGPT and ZiGong are two models used in fraud detection in credit and banking transactions \cite{4a, 28a}. In \cite{17a} credit unions predict fraud and default, while Yan et al. \cite{41a} examine the performance of GPT/BERT-based models on AML, sanctions screening, and suspicious activity.

\item  \textbf{Investment, Trading, and Asset Management}: Walia et al. \cite{18a} apply LLMs to bond yield prediction and extend the use of LLMs in finance. In \cite{27a} LLMs are used for SEC filing interpretation and sentiment-based trading, while Liu et al. \cite{30a} evaluate the performance of FinGPT in robo-advisory and algorithmic trading bots. Other studies test the capabilities of LLMs using Open-FinLLMs in multimodal investment reasoning \cite{35a}. Ni et al. \cite{44a} report that QLoRA outperforms GPT-4 in stock prediction.

\item  \textbf{Taxonomy Building and Transaction Analysis}: In \cite{34a}, LLMs are used to create financial taxonomy and explain transaction data to perform credit analyses, thus strengthening banking behaviors in credit and KYC (Know Your Customer) scenarios.

\item  \textbf{Supply Chain and Sector-Specific Credit Evaluation}: In \cite{38a}, LLMs use bid data and operational descriptions to evaluate China’s green transportation sector SME loans, helping non-specialist analysts understand feasible financing targets and strengthening the accessibility of LLMs.

\item  \textbf{Early Warning Systems and Multidimensional Risk Ratings}: ChatGPT was used for credit risk estimation scenarios where data is incomplete, such as start-ups \cite{39a}.This provides early warning and adaptive credit scoring capabilities for variable domains and applications.

\end{itemize}

\section{Research Gaps and Future Directions (RQ5)}\label{sec:5}

\begin{figure}[h]
    \centering
    \includegraphics[width=0.5\linewidth]{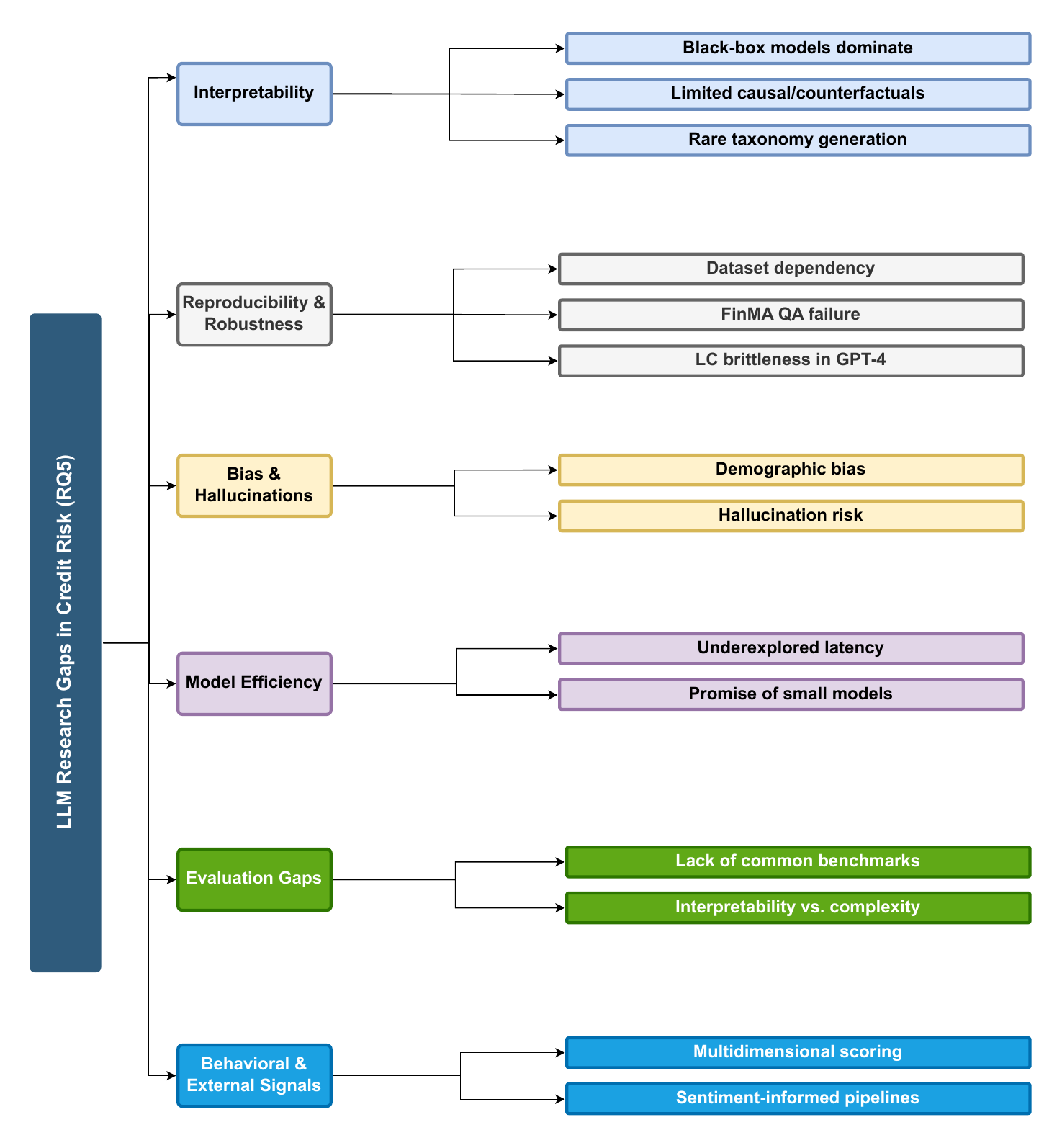}
    \caption{Taxonomy of Research Gaps and Future Directions in LLM-Driven Credit Risk Research. }
    \label{fig:taxonomy_rq5_gaps}
\end{figure}

Although significant advances have been made in LLM-based financial applications (credit risk estimation, etc.), there are still research gaps and limitations that need to be investigated. Figure \ref{fig:taxonomy_rq5_gaps} illustrates these limitations.

\begin{itemize}
    
\item \textbf{Interpretability Gaps}: LLMs applied in finance generally remain black boxes because they rely on post hoc methods (such as SHAP and attention maps) for explainability \cite{5a, 9a}. Causal or counterfactual reasoning is almost absent in LLMs applied in finance, and therefore it is difficult to understand the real decision mechanism of the model. Although recent studies (\cite{13a}) propose counterfactual reasoning, its adoption remains minimal. Furthermore, although research is ongoing on LLMs being able to perform automatic classification (\cite{34a}), they are still not widely used in taxonomy studies.

\item \textbf{Reproducibility and Robustness Limitations}: Few literature studies have performed robustness testing. Most experiments appear to use small or parsimonious datasets \cite{6a, 8a}. Xie et al. \cite{32a} reported that FinMA struggles with reasoning tasks and that models trained with human-like instructions are vulnerable.

\item \textbf{Bias, Fairness, and Hallucination Risks}: The literature reports that LLMs are prone to bias against demographic characteristics such as race, gender, and age \cite{9a, 29a}. Furthermore, LLMs can produce false information, hallucinations, that are very close to the truth, and these remain a major threat to credit risk estimation \cite{10a}. Few frameworks in the literature target fairness and hallucination reduction \cite{41a}.

\item \textbf{Efficiency and Model Scaling}: Very few of the studies consider metrics such as latency, inference cost, and hardware performance \cite{8a, 12a}. Although models like FinLLaMA are calculated considering their small footprint, there is still a research gap in this area \cite{35a}.

\item \textbf{Evaluation Benchmark Deficiencies}: Due to the diversity of datasets and scenarios, it is difficult to directly compare the performance of LLMs, a shortcoming noted in some literature studies (such as the assessment of the trade-off between interpretability and complexity) \cite{27a,59a}.

\item \textbf{Integration of Behavioral and External Signals}: Signals from unstructured data (news, social media content, etc.) can be incorporated into credit scoring models and used for forecasting purposes. Although theoretically potential, behavioral and external signals have not yet been widely integrated.

\end{itemize}

\textbf{Future Directions}: Based on the research gaps described above, the following areas stand out for future researchers:

\begin{itemize}

\item  \textbf{Low-Cost Models}: Compact models such as QLoRA-fine-tuned LLaMA (\cite{44a}) variants can be investigated to develop cost-aware credit scoring models.

\item  \textbf{Evaluation Criteria}: Fairness-focused task-specific evaluation criteria such as context length decay and hallucination robustness can be developed.

\item  \textbf{Fairness and Trust Protocols}: In future research, auditing tools and trust frameworks can be created for consumer finance.

\item  \textbf{Sentiment-Driven Credit Signals}: New studies can be conducted by including concepts such as investor behavior and textual sentiment for loan pricing.

\item  \textbf{Behavioral Personalization}: New LLM-based studies can develop digital banking applications with intent-aware and emotion-aware personalization.

\item  \textbf{Regulatory Consistency and Compliance}: Studies can be conducted on making LLM models compliant with ethical and legal standards in AML applications and anti-bias applications.

\end{itemize}

\section{Conclusions}\label{sec:6}

This paper presents the first systematic review and taxonomy of Large Language Model (LLM) based credit risk assessment approaches. The most relevant 60 peer-reviewed studies published between 2020 and 2025 are analyzed using the PRISMA methodology and a structured taxonomy is presented along four main dimensions: model architectures, data formats, interpretability mechanisms, and application domains. The findings confirm that although only coding and domain-specific FinLLMs are frequently used in this field, recent trends in hybrid pipelines, parameter-efficient tuning (e.g. QLoRA), and multimodal data integration are also used in credit scoring. Although SHAP and LIMA (post-hoc) are prevalent interpretability techniques, there is increasing interest in intrinsic and demand-based explanations. Apart from all these developments, there is still a need for further research in the areas of reproducibility, fairness control, robustness to hallucinations, and standardized assessment. We hope that this systematic review and taxonomy study will be a reference paper for researchers in transparent, reliable and field-compatible LLM-based financial risk modeling.



\bibliographystyle{unsrt}
\bibliography{Ref}

\end{document}